\documentclass[aps,prl,twocolumn,showpacs,superscriptaddress]{revtex4}
\usepackage{graphicx}

\begin{document}
\title{$\Lambda$ and $\overline{\Lambda}$ Production in Central Pb-Pb Collisions
at 40, 80, and 158~A$\cdot$GeV} 

\affiliation{NIKHEF, Amsterdam, The Netherlands.}
\affiliation{Department of Physics, University of Athens, Athens, Greece.}
\affiliation{Comenius University, Bratislava, Slovakia.}
\affiliation{KFKI Research Institute for Particle and Nuclear Physics, Budapest, Hungary.}
\affiliation{MIT, Cambridge, USA.}
\affiliation{Institute of Nuclear Physics, Cracow, Poland.}
\affiliation{Gesellschaft f\"{u}r Schwerionenforschung (GSI), Darmstadt, Germany.}
\affiliation{Joint Institute for Nuclear Research, Dubna, Russia.}
\affiliation{Fachbereich Physik der Universit\"{a}t, Frankfurt, Germany.}
\affiliation{CERN, Geneva, Switzerland.}
\affiliation{University of Houston, Houston, TX, USA.}
\affiliation{Institute of Physics \'Swi{\,e}tokrzyska Academy, Kielce, Poland.}
\affiliation{Fachbereich Physik der Universit\"{a}t, Marburg, Germany.}
\affiliation{Max-Planck-Institut f\"{u}r Physik, Munich, Germany.}
\affiliation{Institute of Particle and Nuclear Physics, Charles University, Prague, Czech Republic.}
\affiliation{Department of Physics, Pusan National University, Pusan, Republic of Korea.}
\affiliation{Nuclear Physics Laboratory, University of Washington, Seattle, WA, USA.}
\affiliation{Atomic Physics Department, Sofia University St. Kliment Ohridski, Sofia, Bulgaria.}
\affiliation{Institute for Nuclear Studies, Warsaw, Poland.}
\affiliation{Institute for Experimental Physics, University of Warsaw, Warsaw, Poland.}
\affiliation{Rudjer Boskovic Institute, Zagreb, Croatia.}

\author{T.~Anticic} 
\affiliation{Rudjer Boskovic Institute, Zagreb, Croatia.}
\author{B.~Baatar}
\affiliation{Joint Institute for Nuclear Research, Dubna, Russia.}
\author{D.~Barna}
\affiliation{KFKI Research Institute for Particle and Nuclear Physics, Budapest, Hungary.}
\author{J.~Bartke}
\affiliation{Institute of Nuclear Physics, Cracow, Poland.}
\author{M.~Behler}
\affiliation{Fachbereich Physik der Universit\"{a}t, Marburg, Germany.}
\author{L.~Betev}
\affiliation{Fachbereich Physik der Universit\"{a}t, Frankfurt, Germany.}
\author{H.~Bia{\l}\-kowska} 
\affiliation{Institute for Nuclear Studies, Warsaw, Poland.}
\author{A.~Billmeier}
\affiliation{Fachbereich Physik der Universit\"{a}t, Frankfurt, Germany.}
\author{C.~Blume}
\affiliation{Gesellschaft f\"{u}r Schwerionenforschung (GSI), Darmstadt, Germany.}
\author{B.~Boimska}
\affiliation{Institute for Nuclear Studies, Warsaw, Poland.}
\author{M.~Botje}
\affiliation{NIKHEF, Amsterdam, The Netherlands.}
\author{J.~Bracinik}
\affiliation{Comenius University, Bratislava, Slovakia.}
\author{R.~Bramm}
\affiliation{Fachbereich Physik der Universit\"{a}t, Frankfurt, Germany.}
\author{R.~Brun}
\affiliation{CERN, Geneva, Switzerland.}
\author{P.~Bun\v{c}i\'{c}}
\affiliation{CERN, Geneva, Switzerland.}
\affiliation{Fachbereich Physik der Universit\"{a}t, Frankfurt, Germany.}
\author{V.~Cerny}
\affiliation{Comenius University, Bratislava, Slovakia.}
\author{P.~Christakoglou}
\affiliation{Department of Physics, University of Athens, Athens, Greece.}
\author{O.~Chvala}
\affiliation{Institute of Particle and Nuclear Physics, Charles University, Prague, 
Czech Republic.}
\author{J.G.~Cramer}
\affiliation{Nuclear Physics Laboratory, University of Washington, Seattle, WA, USA.}
\author{P.~Csat\'{o}} 
\affiliation{KFKI Research Institute for Particle and Nuclear Physics, Budapest, Hungary.}
\author{N.~Darmenov}
\affiliation{Atomic Physics Department, Sofia University St. Kliment Ohridski, Sofia, Bulgaria.}
\author{A.~Dimitrov}
\affiliation{Atomic Physics Department, Sofia University St. Kliment Ohridski, Sofia, Bulgaria.}
\author{P.~Dinkelaker}
\affiliation{Fachbereich Physik der Universit\"{a}t, Frankfurt, Germany.}
\author{V.~Eckardt}
\affiliation{Max-Planck-Institut f\"{u}r Physik, Munich, Germany.}
\author{P.~Filip}
\affiliation{Max-Planck-Institut f\"{u}r Physik, Munich, Germany.}
\author{D.~Flierl}
\affiliation{Fachbereich Physik der Universit\"{a}t, Frankfurt, Germany.}
\author{Z.~Fodor}
\affiliation{KFKI Research Institute for Particle and Nuclear Physics, Budapest, Hungary.}
\author{P.~Foka}
\affiliation{Gesellschaft f\"{u}r Schwerionenforschung (GSI), Darmstadt, Germany.}
\author{P.~Freund}
\affiliation{Max-Planck-Institut f\"{u}r Physik, Munich, Germany.}
\author{V.~Friese}
\affiliation{Gesellschaft f\"{u}r Schwerionenforschung (GSI), Darmstadt, Germany.}
\affiliation{Fachbereich Physik der Universit\"{a}t, Marburg, Germany.}
\author{J.~G\'{a}l}
\affiliation{KFKI Research Institute for Particle and Nuclear Physics, Budapest, Hungary.}
\author{M.~Ga\'zdzicki}
\affiliation{Fachbereich Physik der Universit\"{a}t, Frankfurt, Germany.}
\author{G.~Georgopoulos}
\affiliation{Department of Physics, University of Athens, Athens, Greece.}
\author{E.~G{\l}adysz}
\affiliation{Institute of Nuclear Physics, Cracow, Poland.}
\author{S.~Hegyi}
\affiliation{KFKI Research Institute for Particle and Nuclear Physics, Budapest, Hungary.}
\author{C.~H\"{o}hne}
\affiliation{Fachbereich Physik der Universit\"{a}t, Marburg, Germany.}
\author{K.~Kadija}
\affiliation{CERN, Geneva, Switzerland.}
\affiliation{Rudjer Boskovic Institute, Zagreb, Croatia.}
\author{A.~Karev}
\affiliation{Max-Planck-Institut f\"{u}r Physik, Munich, Germany.}
\author{V.I.~Kolesnikov}
\affiliation{Joint Institute for Nuclear Research, Dubna, Russia.}
\author{T.~Kollegger}
\affiliation{Fachbereich Physik der Universit\"{a}t, Frankfurt, Germany.}
\author{E.~Kornas}
\affiliation{Institute of Nuclear Physics, Cracow, Poland.}
\author{R.~Korus}
\affiliation{Institute of Physics \'Swi{\,e}tokrzyska Academy, Kielce, Poland.}
\author{M.~Kowalski}
\affiliation{Institute of Nuclear Physics, Cracow, Poland.}
\author{I.~Kraus}
\affiliation{Gesellschaft f\"{u}r Schwerionenforschung (GSI), Darmstadt, Germany.}
\author{M.~Kreps}
\affiliation{Comenius University, Bratislava, Slovakia.}
\author{M.~van~Leeuwen}
\affiliation{NIKHEF, Amsterdam, The Netherlands.}
\author{P.~L\'{e}vai}
\affiliation{KFKI Research Institute for Particle and Nuclear Physics, Budapest, Hungary.}
\author{L.~Litov}
\affiliation{Atomic Physics Department, Sofia University St. Kliment Ohridski, Sofia, Bulgaria.}
\author{M.~Makariev}
\affiliation{Atomic Physics Department, Sofia University St. Kliment Ohridski, Sofia, Bulgaria.}
\author{A.I.~Malakhov}
\affiliation{Joint Institute for Nuclear Research, Dubna, Russia.}
\author{C.~Markert}
\affiliation{Gesellschaft f\"{u}r Schwerionenforschung (GSI), Darmstadt, Germany.}
\author{M.~Mateev}
\affiliation{Atomic Physics Department, Sofia University St. Kliment Ohridski, Sofia, Bulgaria.}
\author{B.W.~Mayes}
\affiliation{University of Houston, Houston, TX, USA.}
\author{G.L.~Melkumov}
\affiliation{Joint Institute for Nuclear Research, Dubna, Russia.}
\author{C.~Meurer}
\affiliation{Fachbereich Physik der Universit\"{a}t, Frankfurt, Germany.}
\author{A.~Mischke}
\affiliation{Gesellschaft f\"{u}r Schwerionenforschung (GSI), Darmstadt, Germany.}
\author{M.~Mitrovski}
\affiliation{Fachbereich Physik der Universit\"{a}t, Frankfurt, Germany.}
\author{J.~Moln\'{a}r}
\affiliation{KFKI Research Institute for Particle and Nuclear Physics,
Budapest, Hungary.}
\author{St.~Mr\'owczy\'nski}
\affiliation{Institute of Physics \'Swi{\,e}tokrzyska Academy, Kielce, Poland.}
\author{G.~P\'{a}lla}
\affiliation{KFKI Research Institute for Particle and Nuclear Physics, Budapest, Hungary.}
\author{A.D.~Panagiotou}
\affiliation{Department of Physics, University of Athens, Athens, Greece.}
\author{D.~Panayotov}
\affiliation{Atomic Physics Department, Sofia University St. Kliment Ohridski, Sofia, Bulgaria.}
\author{K.~Perl}
\affiliation{Institute for Experimental Physics, University of Warsaw, Warsaw, Poland.}
\author{A.~Petridis}
\affiliation{Department of Physics, University of Athens, Athens, Greece.}
\author{M.~Pikna}
\affiliation{Comenius University, Bratislava, Slovakia.}
\author{L.~Pinsky}
\affiliation{University of Houston, Houston, TX, USA.}
\author{F.~P\"{u}hlhofer}
\affiliation{Fachbereich Physik der Universit\"{a}t, Marburg, Germany.}
\author{J.G.~Reid}
\affiliation{Nuclear Physics Laboratory, University of Washington, Seattle, WA, USA.}
\author{R.~Renfordt}
\affiliation{Fachbereich Physik der Universit\"{a}t, Frankfurt, Germany.}
\author{W.~Retyk}
\affiliation{Institute for Experimental Physics, University of Warsaw, Warsaw, Poland.}
\author{C.~Roland}
\affiliation{MIT, Cambridge, USA.}
\author{G.~Roland}
\affiliation{MIT, Cambridge, USA.}
\author{M. Rybczy\'nski}
\affiliation{Institute of Physics \'Swi{\,e}tokrzyska Academy, Kielce, Poland.}
\author{A.~Rybicki}
\affiliation{Institute of Nuclear Physics, Cracow, Poland.}
\author{A.~Sandoval}
\affiliation{Gesellschaft f\"{u}r Schwerionenforschung (GSI), Darmstadt, Germany.}
\author{H.~Sann}
\affiliation{Gesellschaft f\"{u}r Schwerionenforschung (GSI), Darmstadt, Germany.}
\author{N.~Schmitz}
\affiliation{Max-Planck-Institut f\"{u}r Physik, Munich, Germany.}
\author{P.~Seyboth}
\affiliation{Max-Planck-Institut f\"{u}r Physik, Munich, Germany.}
\author{F.~Sikl\'{e}r}
\affiliation{KFKI Research Institute for Particle and Nuclear Physics, Budapest, Hungary.}
\author{B.~Sitar}
\affiliation{Comenius University, Bratislava, Slovakia.}
\author{E.~Skrzypczak}
\affiliation{Institute for Experimental Physics, University of Warsaw, Warsaw, Poland.}
\author{G.~Stefanek}
\affiliation{Institute of Physics \'Swi{\,e}tokrzyska Academy, Kielce, Poland.}
\author{R.~Stock}
\affiliation{Fachbereich Physik der Universit\"{a}t, Frankfurt, Germany.}
\author{H.~Str\"{o}bele}
\affiliation{Fachbereich Physik der Universit\"{a}t, Frankfurt, Germany.}
\author{T.~Susa}
\affiliation{Rudjer Boskovic Institute, Zagreb, Croatia.}
\author{I.~Szentp\'{e}tery}
\affiliation{KFKI Research Institute for Particle and Nuclear Physics, Budapest, Hungary.}
\author{J.~Sziklai}
\affiliation{KFKI Research Institute for Particle and Nuclear Physics, Budapest, Hungary.}
\author{T.A.~Trainor}
\affiliation{Nuclear Physics Laboratory, University of Washington, Seattle, WA, USA.}
\author{D.~Varga}
\affiliation{KFKI Research Institute for Particle and Nuclear Physics, Budapest, Hungary.}
\author{M.~Vassiliou}
\affiliation{Department of Physics, University of Athens, Athens, Greece.}
\author{G.I.~Veres}
\affiliation{KFKI Research Institute for Particle and Nuclear Physics, Budapest, Hungary.}
\author{G.~Vesztergombi}
\affiliation{KFKI Research Institute for Particle and Nuclear Physics, Budapest, Hungary.}
\author{D.~Vrani\'{c}}
\affiliation{Gesellschaft f\"{u}r Schwerionenforschung (GSI), Darmstadt, Germany.}
\author{A.~Wetzler}
\affiliation{Fachbereich Physik der Universit\"{a}t, Frankfurt, Germany.}
\author{Z.~W{\l}odarczyk}
\affiliation{Institute of Physics \'Swi{\,e}tokrzyska Academy, Kielce, Poland.}
\author{I.K.~Yoo}
\affiliation{Department of Physics, Pusan National University, Pusan, Republic of Korea.}
\author{J.~Zaranek}
\affiliation{Fachbereich Physik der Universit\"{a}t, Frankfurt, Germany.}
\author{J.~Zim\'{a}nyi}
\affiliation{KFKI Research Institute for Particle and Nuclear Physics, Budapest, Hungary.}
\collaboration{The NA49 Collaboration}
\noaffiliation
\date{\today}

\begin{abstract}
Production of Lambda and Antilambda hyperons was measured in central
Pb-Pb collisions at 40, 80, and 158~A$\cdot$GeV beam energy on a fixed target. 
Transverse mass spectra and rapidity distributions are given for all 
three energies.
The $\Lambda/\pi$ ratio at mid-rapidity and in full phase space shows a 
pronounced maximum between the highest AGS and 40~A$\cdot$GeV 
SPS energies, whereas the $\overline{\Lambda}/\pi$ ratio exhibits a 
monotonic increase. 
\end{abstract}
%
\maketitle

Relativistic nucleus-nucleus collisions allow the investigation of
hadronic matter at high temperatures and densities. One of the
crucial features of nuclear collisions is the relative increase of
strange particle production as compared to elementary
hadron-hadron collisions. 
Systematic studies of hadron production in central A-A collisions have shown 
that this strangeness enhancement is not specific to the top SPS~\cite{Bar90,Alb94}
and RHIC~\cite{STAR02,Adc02} energy range, $\sqrt{s_{\rm NN}}$ from 18 
to 200 GeV, but also occurs at the much lower AGS~\cite{Ahl99} and 
SIS~\cite{Cley00} energies ($\sqrt{s_{\rm NN}} <$ 6~GeV). 
A theoretical view of the strangeness systematics was recently
obtained when applying the Hagedorn statistical hadronization
model~\cite{Hag79} in its grand canonical form. 
Contrary to the earlier analysis~\cite{Raf821} 
which primarily focused on the energy density (assumed to be above the critical 
deconfinement phase transition energy density of about 1~GeV/fm$^3$ 
predicted by lattice QCD~\cite{Kar02}) the crucial parameter in the statistical
model is the large coherent volume of the high density
fireball~\cite{TouRed202} which is characteristic of central nucleus-nucleus 
collisions. The severe constraints of local strangeness conservation, 
characteristic of small volume elementary collisions, disappear, leading to an 
increase of the ratio of strange to non-strange hadron yields. 

It was suggested~\cite{GazGor99} that the onset of deconfinement 
should cause a non-monotonic energy dependence of the total strangeness to 
pion ratio. 
This effect was recently observed in NA49 data on the energy dependence of
kaon and pion production in central Pb-Pb collisions~\cite{KPI02,Fri03} 
where a sharp
maximum of the K$^+/\pi^+$ ratio is seen at 30~A$\cdot$GeV beam energy.
To obtain an estimate of the energy
dependence of total strangeness production and to study how the strange quarks
and anti-strange quarks are distributed among the relevant hadronic
species, it is important to complement the data of Refs.~\cite{KPI02,Fri03} 
on K$^+$ and K$^-$ yields by data on both $\Lambda$ and $\overline{\Lambda}$ 
production.

In this letter we present measurements of $\Lambda$ and $\overline{\Lambda}$ 
production in central Pb-Pb collisions at 40, 80, and 158~GeV per nucleon 
over a wide range in rapidity  
($|y|\le1.6$, where $y$ is the rapidity in the cms) 
and transverse mass $m_{\rm T}$ ($0\le(m_{\rm T}-m_0)\le1.6$ GeV/c$^2$,
where $m_0$ is the $\Lambda$ mass). 
Preliminary analyses have been reported in Refs.~\cite{Mis02,Mis03}.

The NA49 detector is a large acceptance hadron spectrometer~\cite{NIM99}.  
Tracking and particle identification by measuring the
specific energy loss (d$E$/d$x$) is performed by two Time Projection
Chambers (Vertex-TPCs), {\mbox located} inside two vertex magnets, and 
two large volume TPCs (Main-TPCs) situated downstream of the magnets 
symmetrically to the beam line. 
The relative d$E$/d$x$ resolution is 4$\%$ and the momentum resolution  
$\sigma(p)/p^2$ = 0.3 $\cdot$ $10^{-4}$ (GeV/c)$^{-1}$. 
Centrality selection is based on a measurement of the energy 
deposited in a forward calorimeter by the projectile spectator nucleons.
For the present analysis, the 7.2$\%$ most
central interactions at 40 and 80~A$\cdot$GeV were selected.
Using the Glauber model to convert a cross section fraction into the 
number of wounded nucleons ($\langle {\rm N}_{\rm W}\rangle$) 
per event this corresponds, on average, to 
$\langle {\rm N}_{\rm W}\rangle$ = 349~\cite{KPI02}.
For 158~A$\cdot$GeV the 10$\%$ most central events were selected
($\langle {\rm N}_{\rm W}\rangle$ = 335).
About 400\hspace{0.5mm}000 events were analyzed for 40 and
158~A$\cdot$GeV each and 300\hspace{0.5mm}000 events for 80~A$\cdot$GeV.  

\begin{figure}[t]
\begin{minipage}[b]{4.cm}
 \includegraphics[width=3.6cm]{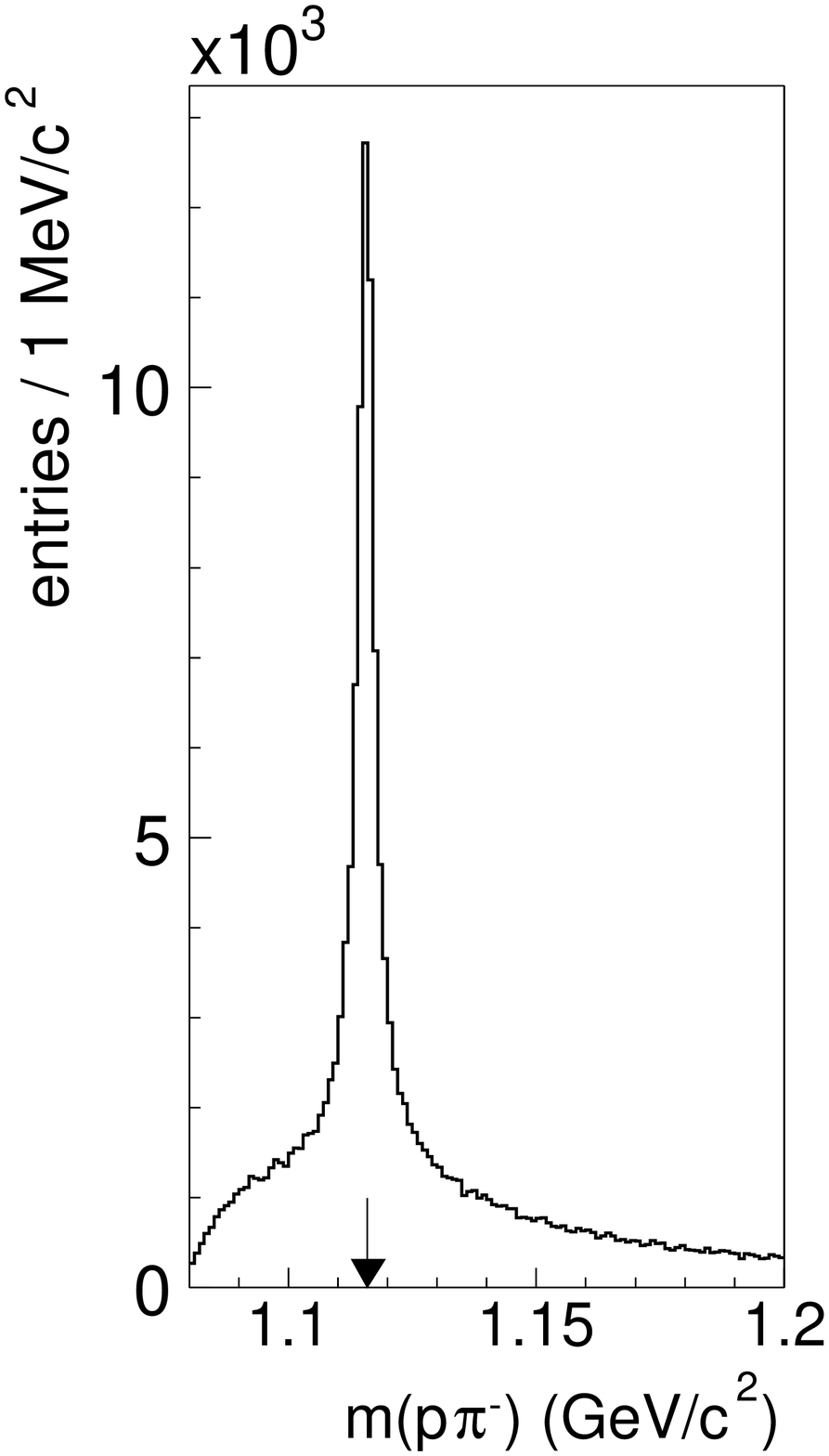}
\end{minipage} 
%
%
\begin{minipage}[b]{4.cm}
 \includegraphics[width=3.6cm]{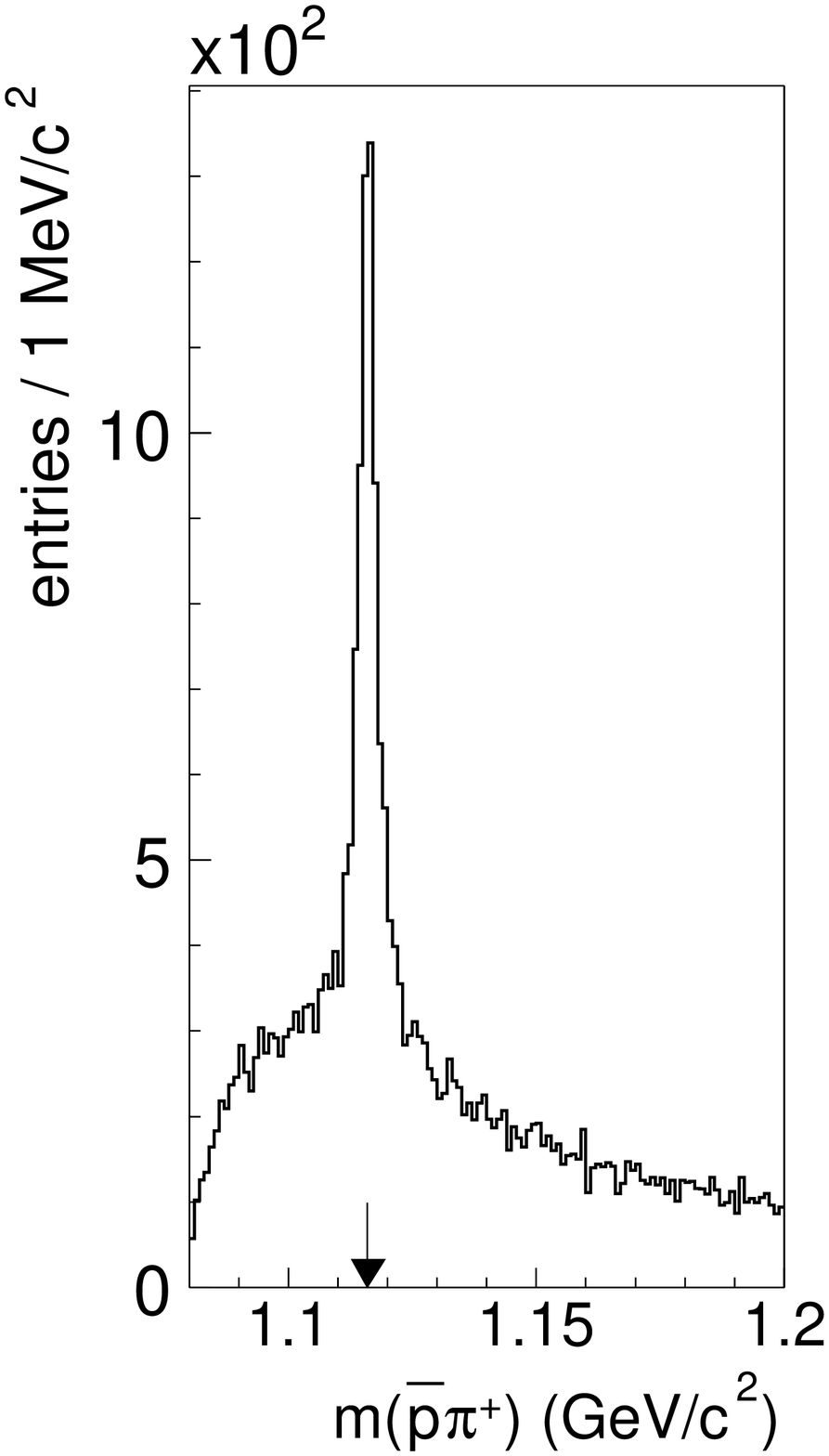}
\end{minipage} 
\vspace{-0.35cm}
\caption{\protect \footnotesize 
Invariant mass distribution of $\Lambda$ (left) and $\overline{\Lambda}$
(right) candidates in central Pb-Pb reactions at 158~GeV per nucleon. 
The $\Lambda$ ($\overline{\Lambda}$) PDG mass (1.115~GeV/c$^2$) is
indicated by the arrows.}  
\label{fig1}
\end{figure}

\begin{figure}[t]
\begin{minipage}[b]{4.25cm}
 \begin{center}
 \includegraphics[width=4.25cm]{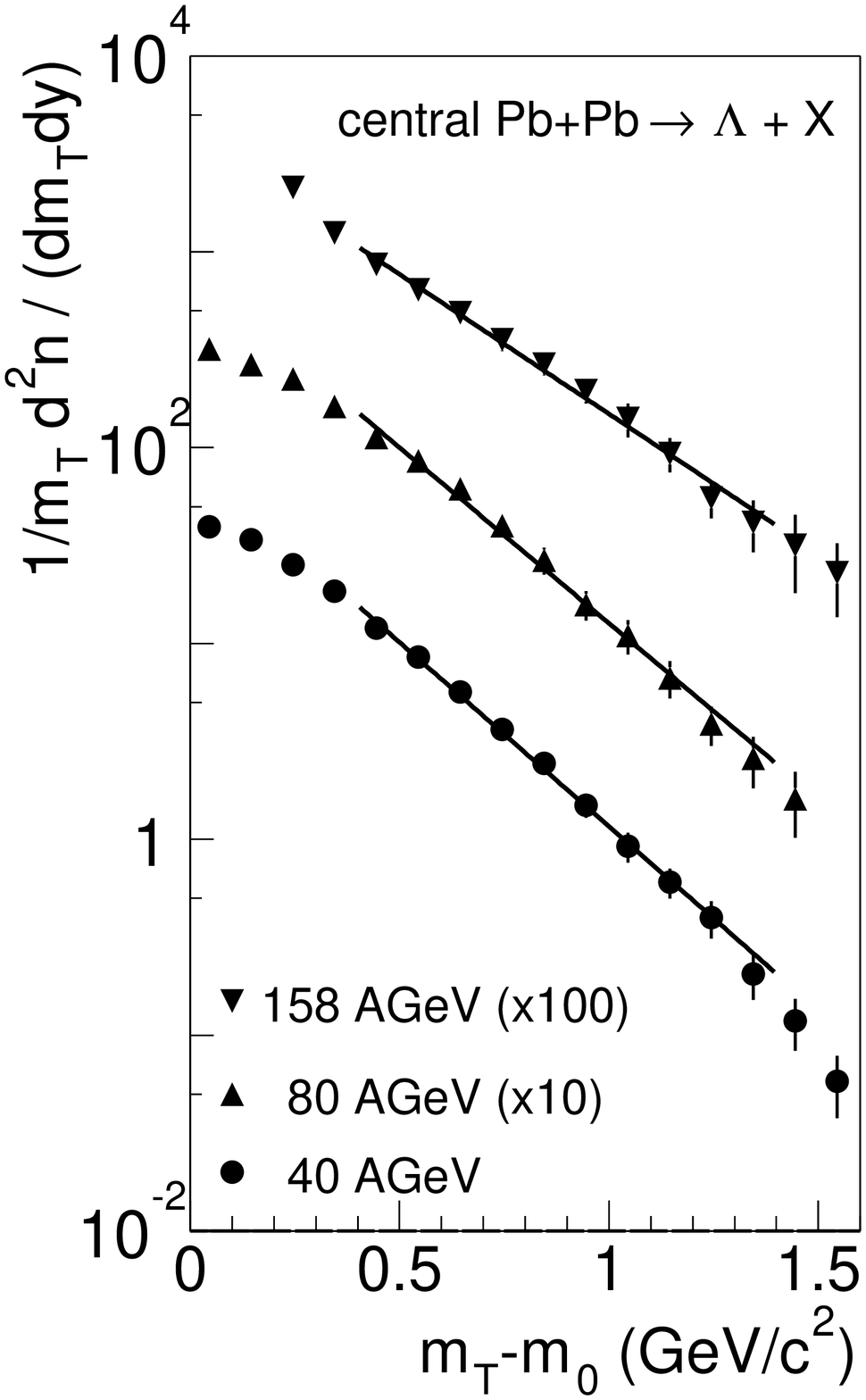}
 \end{center}
\end{minipage} 
%
%
\begin{minipage}[b]{4.25cm}
 \includegraphics[width=4.3cm]{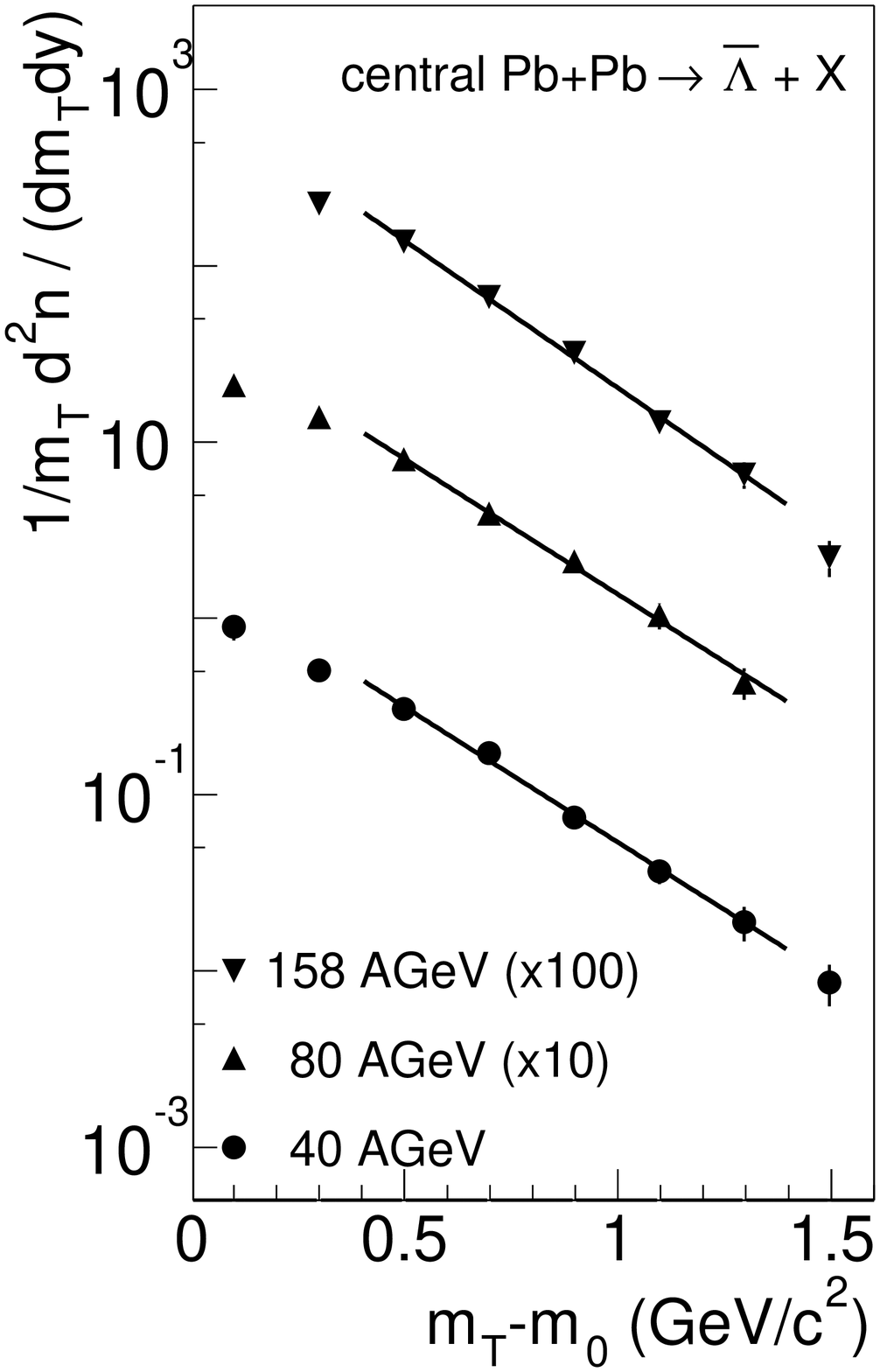}
\end{minipage} 
\vspace{-0.8cm}
\caption{\protect \footnotesize 
Transverse mass spectra of $\Lambda$ and $\overline{\Lambda}$ at mid-rapidity 
($|y| \le 0.4$).
The full lines are exponential fits described in the text.}  
\label{fig2}
\end{figure}

$\Lambda$ and $\overline{\Lambda}$ hyperons are identified
by reconstructing their decay topologies \mbox{ $\Lambda \rightarrow 
p+\pi^{-}$} and $\overline{\Lambda} \rightarrow \overline{p}+\pi^{+}$, 
respectively (branching ratio 63.9$\%$). 
Candidates were found by forming pairs from all measured positively
and negatively charged particles requiring a distance of closest approach
between the two trajectories of less than 1~cm at any point before
reaching the target plane.
To reduce the combinatorial background from random pairs a set 
of quality cuts~\cite{myphd} was imposed on the position of the
secondary vertex (at least 30~cm downstream from the target and
outside the sensitive volume of the TPCs), on the impact parameter of 
the parent and the daughter tracks in the target plane, and on the number 
of points measured in the TPC.
At 158~A$\cdot$GeV an additional geometric quality cut was applied, which 
excludes particles in the high track density region from the 
analysis~\cite{myphd}.
This results in a decreased acceptance for $\Lambda$ and $\overline{\Lambda}$
at low transverse mass.
For each $\Lambda$ ($\overline{\Lambda}$) candidate the invariant mass 
was calculated assuming that the positive track is a proton ($\pi^+$) 
and the negative track is a $\pi^-$ ($\overline{\rm p}$).
To enrich the decay protons (anti-protons) a cut on the specific energy 
loss d$E$/d$x$ of the positive (negative) tracks of $\pm 4\sigma$ from 
the expected mean value was applied.
In Fig.~\ref{fig1}, the resulting invariant mass distributions of
(p$\pi^{-}$) and ($\overline{\rm p}\pi^{+}$) pairs at 158 A$\cdot$GeV 
are shown. Clear signals are observed. 
The peak positions are in agreement with the nominal value
of the Lambda hyperon mass~\cite{PDG98}. 
The mass resolution ($\sigma_{\rm m}$) is about 2~MeV/c$^2$ at all 
three energies.  
The background was subtracted using a third-order polynomial.
Corrections for geometrical acceptance, branching ratio and
tracking efficiency were calculated bin by bin in rapidity and
transverse momentum using GEANT 3.21 for detector simulation and 
dedicated NA49 simulation software~\cite{myphd}.
The systematic errors were estimated by varying the quality cuts and
by analyzing selected subvolumes of the TPCs and were found to be
smaller than 9$\%$.  
Corrections for feed-down from weak decays (mostly $\Xi^-$, $\Xi^0$
and their anti-particles) are not applied and were estimated to be about 
(6$\pm$3)\% for $\Lambda$ and (12$\pm$6)\% for $\overline{\Lambda}$.
A detailed study shows a weak $p_{\rm T}$ and rapidity dependence.
In the following $\Lambda$ ($\overline{\Lambda}$) include those from 
electro-magnetic $\Sigma^{0}$ ($\overline{\Sigma}^{0}$) decays.
The transverse mass distributions 
at mid-rapidity ($|y| \le 0.4$) are shown in Fig.~\ref{fig2}. 
All spectra are fitted by an exponential function in $m_{\rm T}$:
\begin{eqnarray*}
 \frac{1}{m_{\rm T}} \frac{{\rm d}^2n}{{\rm d}m_{\rm T} {\rm d}y}
\propto \exp\left(-\frac{m_{\rm T}}{T}\right), 
\label{eq1}
\end{eqnarray*}
\noindent
where $T$ is the inverse slope parameter.  
The fitted range (in $m_{\rm T}-m_0$) is 0.4--1.4~GeV/c$^2$. 
The results are summarized in Tab.~\ref{tab1}.
In this fit region, the $\Lambda$ inverse slope parameter $T$
increases with collision energy.
The deviations at low transverse mass, seen in Fig.~\ref{fig2} for 40 
and 80~A$\cdot$GeV, and the convex shape of the spectra, indicate the 
effect of transverse flow~\cite{Schn93}.
\begin{table}[b]
\begin{center}
\begin{tabular}{c c c c}
 \hline
 \hline
 & \hspace{2mm} 40 A$\cdot$GeV \hspace{2mm} & \hspace{2mm} 80 A$\cdot$GeV \hspace{2mm} & \hspace{2mm} 158 A$\cdot$GeV \hspace{2mm} \\
 \hline
 $T$($\Lambda$) (MeV)             & 231$\pm$8$\pm$20 & 241$\pm$10$\pm$22 & 304$\pm$16$\pm$23 \\  
 $T$($\overline{\Lambda}$) (MeV)  & 283$\pm$16$\pm$20 & 283$\pm$16$\pm$22 & 261$\pm$12$\pm$23 \\  
 d$n$/d$y$ ($\Lambda$)            & 15.3$\pm$0.6$\pm$1.0    & 13.5$\pm$0.7$\pm$1.0   & 10.9$\pm$1.0$\pm$1.3\\  
 d$n$/d$y$ ($\overline{\Lambda}$) & 0.42$\pm$0.04$\pm$0.04  & 1.06$\pm$0.08$\pm$0.1  & 1.62$\pm$0.16$\pm$0.2\\  
 $\sigma$ ($\Lambda$)             & 1.16$\pm$0.06 & - & - \\
 $\sigma$ ($\overline{\Lambda}$)  & 0.71$\pm$0.05 & 0.85$\pm$0.13 & 0.95$\pm$0.05\\

 $\langle\Lambda\rangle$            & 45.6$\pm$1.9$\pm$3.4   & 47.4$\pm$2.8$\pm$3.5  & 44.1$\pm$3.2$\pm$5.0 \\ 
 $\langle\overline{\Lambda}\rangle$ & 0.74$\pm$0.04$\pm$0.06 & 2.26$\pm$0.25$\pm$0.2 & 3.87$\pm$0.18$\pm$0.4 \\ 
\hline
\hline
\end{tabular}    
\caption{\label{tab1} \footnotesize 
Inverse slope parameter $T$ of the transverse mass spectra,
fitted in the $(m_{\rm T}-m_0)$ range of 0.4--1.4~GeV/c$^2$,
and the
rapidity density d$n$/d$y$, both at mid-rapidity ($|y| \le 0.4$), the
width $\sigma$ of the Gaussian fits to the rapidity distribution and 
the total multiplicity for $\Lambda$ and $\overline{\Lambda}$ hyperons. 
The first error is statistical, the second systematic.
Feed-down from weak decays are estimated to be about (6$\pm$3)\% for $\Lambda$ 
and (12$\pm$6)\% for $\overline{\Lambda}$.}  
\end{center}
\end{table}
\begin{figure}[b]
 \includegraphics[width=9.cm]{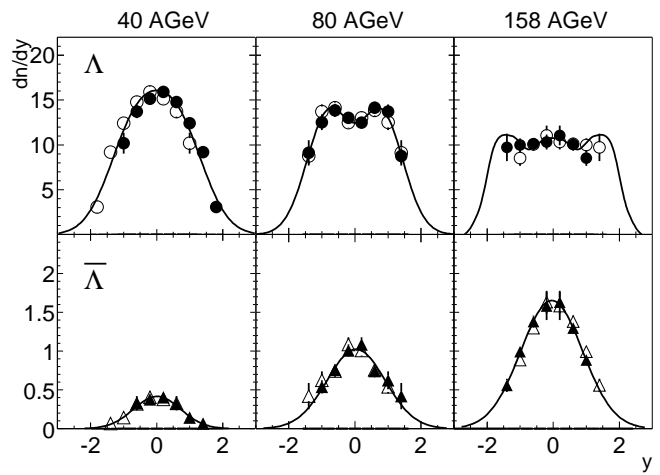}
\vspace{-0.75cm}
\caption{\protect \footnotesize 
Rapidity distribution of $\Lambda$ (top) and $\overline{\Lambda}$
(bottom) at 40, 80, and 158~A$\cdot$GeV beam energy (full symbols).
The open symbols show the measured points, reflected with respect to $y$ = 0.
The errors are statistical only. 
The full lines represent fits to the data which were used to obtain total yields.}  
\label{fig3}
\end{figure}
Rapidity distributions are obtained by integrating the measured
$p_{\rm T}$ spectra and by extra\-polation into unmeasured regions.
At 40 and 80~A$\cdot$GeV the acceptance covers the full transverse 
momentum range down to $p_{\rm T}$~=~0.
At 158~A$\cdot$GeV the $p_{\rm T}$ integration
was started at $p_{\rm T}$~=~0.6~GeV/c. The extrapolation to the full
$p_{\rm T}$ range was performed by multiplying with factors 1.41 ($\Lambda$)
and 1.35 ($\overline{\Lambda}$), which were derived from the 80~A$\cdot$GeV
$p_{\rm T}$ spectra. 
Using the fitted exponential functions or a combined fit of a blast-wave
model to the NA49 particle spectra at 158~A$\cdot$GeV~\cite{Fri03}
would results in 5-10\% different extra\-polation factors.
This uncertainty is included in the estimated systematic errors.
The resulting rapidity distributions of $\Lambda$ and $\overline{\Lambda}$ 
hyperons are compared in Fig.~\ref{fig3}.
The $\Lambda$ rapidity distribution at 40~A$\cdot$GeV is peaked
at mid-rapidity whereas this distribution becomes broader and flatter
with increasing energy.
Since $\Lambda$ hyperons carry a significant fraction of the total
net baryon number their rapidity distribution reflects the overall
net baryon number distribution which is not peaked at mid-rapidity
due to incomplete stopping of the incoming nucleons at top SPS energy.
The same behaviour was observed for 
the $y$ distribution of
protons in central Pb-Pb collisions at this energy~\cite{Stop99}.

The $\Lambda$ rapidity density at mid-rapidity (cf. Tab.~\ref{tab1}) 
decreases with increasing energy, whereas it increases for 
$\overline{\Lambda}$ hyperons. 
The inverse slope parameter and the mid-rapidity density at 158~A$\cdot$GeV 
agree with previous data from WA97~\cite{WA9799}.
The NA45 collaboration has measured $\Lambda$ at
40~A$\cdot$GeV~\cite{Schm02}. 
Their extracted slope parameter is 40 MeV higher than the present result, 
but compatible within errors. 
Their mid-rapidity density (d$n$/d$y = 11.8\pm2$) also agrees within errors. 

The total multiplicities are obtained by integrating the
rapidity spectra with extrapolations into unmeasured regions
using Gaussian fits for the $\overline{\Lambda}$ at all three energies
and for the $\Lambda$ at 40 A$\cdot$GeV. A double Gaussian fit
is used for the $\Lambda$ at 80 A$\cdot$GeV. 
For the $\Lambda$ at 158~A$\cdot$GeV an
extrapolation is made using an average of the tails of the net-proton 
distribution at 158~A$\cdot$GeV~\cite{Stop99} and the $\Lambda$ rapidity 
distribution in central S-S collisions~\cite{Alb94}.   
The multiplicities in full phase space are summarized in
Tab.~\ref{tab1}.

\begin{figure}[t!]
 \includegraphics[width=7.7cm]{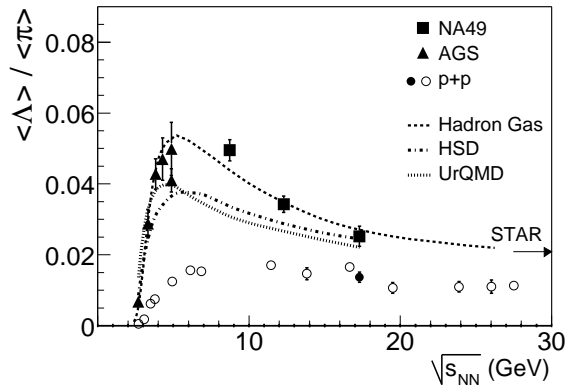}
\vspace{-0.35cm}
\caption{\protect \footnotesize 
The $\langle\Lambda\rangle/\langle\pi\rangle$ ratio in full phase space
versus energy from NA49 (squares), AGS~\cite{Pin02,Alb02,Bec01} (triangles) 
and p-p reactions (Filled circle from NA49 and open circles from 
Refs.~\cite{Gaz96,Gaz95}). 
The STAR measurements at mid-rapidity~\cite{STAR} are indicated by the arrow.
The curves show predictions from the Hadron-gas model~\cite{pbm02} 
(dashed), UrQMD~\cite{Web302} (dotted) and HSD~\cite{Web302} (dash-dotted).}
\label{fig4}
\end{figure}

\begin{figure}[t]
 \includegraphics[width=7.7cm]{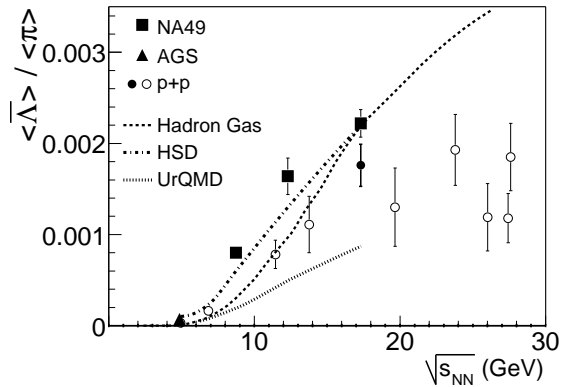}
\vspace{-0.35cm}
\caption{\protect \footnotesize 
The $\langle\overline{\Lambda}\rangle/\langle\pi\rangle$ ratio in full 
phase space versus energy from NA49 (squares), AGS~\cite{Bac01} (triangles) 
and p-p reactions 
(Filled circle from NA49 and open circles from Refs.~\cite{Gaz96,Gaz95}).
The curves show predictions from the Hadron-gas model~\cite{pbm02} 
(dashed), UrQMD~\cite{Web302} (dotted) and HSD~\cite{Web202,Cass02} 
(dash-dotted).}
\label{fig5}
\end{figure}

Remarkably, the $\Lambda$ multiplicity shows no significant change
between 40 and 160~A$\cdot$GeV
whereas $\overline{\Lambda}$ production grows by a factor of about 6.
To compare these results with hyperon production in p-p collisions
the yield is normalized to the mean pion multi\-plicity
$\langle\pi\rangle = 1.5 (\langle\pi^+\rangle + \langle\pi^-\rangle)$,
using measurements from NA49~\cite{KPI02}, E895~\cite{Kla01}, and 
E802~\cite{Ahl198}.    
The energy dependence of the 
$\langle\Lambda\rangle/\langle\pi\rangle$ and 
$\langle\overline{\Lambda}\rangle/\langle\pi\rangle$ 
ratio is given as a function of collision energy $\sqrt{s_{\rm NN}}$ 
in Figs.~\ref{fig4} and~\ref{fig5}.
The $\langle\Lambda\rangle/\langle\pi\rangle$ ratio steeply increases
at AGS energies~\cite{Pin02,Alb02,Bec01} whereas it decreases gradually
at SPS energies. 
The STAR measurement at $\sqrt{s_{\rm NN}} = 130$~GeV~\cite{STAR}
follows this trend. 
The $\langle\overline{\Lambda}\rangle/\langle\pi\rangle$ ratio, however,
shows a monotonic increase up to RHIC energies~\cite{STAR} 
without significant structure.
Qualitatively, the maximum in the $\langle\Lambda\rangle/\langle\pi\rangle$ 
ratio can be understood to arise from the interplay
of the opening of the threshold of $\Lambda$-K associate production and
the rapidly decreasing net baryon density in the produced fireball.
In contrast, $\overline{\Lambda}$ production is not sensitive to the net 
baryon density and shows a continuous threshold increase.
For elementary p-p interactions~\cite{Gaz96,Gaz95} the 
$\langle\Lambda\rangle/\langle\pi\rangle$ ratio shows an increase up to 
$\sqrt{s_{\rm NN}} = 5$~GeV followed by a saturation at higher energies 
(cf. Fig.~\ref{fig4}) with a similar trend observed in the
$\langle\overline{\Lambda}\rangle/\langle\pi\rangle$ ratio 
(cf. Fig.~\ref{fig5}). It is seen that both ratios are, at all energies, 
significantly below the A-A results.
Consequently, $\Lambda$ and $\overline{\Lambda}$ production in A-A
collisions cannot be understood as a superposition of nucleon-nucleon
interactions.
The observed strangeness enhancement
is expected from the statistical hadronization 
model~\cite{TouRed202,pbm02,MRTqm02} which explains it as the
fading-away of canonical strangeness suppression, characterizing
the comparatively small fireball volume in p-p collisions.

Turning to the $\sqrt{s}$-dependence of strange to non-strange
yield ratios we note, first, a close correspondence between the
present $\Lambda$, $\overline{\Lambda}$ hyperon data and the
K$^+$, K$^-$ data previously obtained by NA49~\cite{KPI02} for 
central Pb-Pb collisions at 40, 80, and 158~A$\cdot$GeV. Both the 
$\langle\Lambda\rangle/\langle\pi\rangle$ and
$\langle{\rm K}^+\rangle/\langle\pi\rangle$
ratios exhibit a distinct peak occurring between a
steep rise toward top AGS energies, and a smooth fall-off from 
40~A$\cdot$GeV onward to RHIC energy. On the contrary both
$\langle\overline{\Lambda}\rangle/\langle\pi\rangle$ and
$\langle{\rm K}^-\rangle/\langle\pi\rangle$ ascend
monotonically toward RHIC energy. The latter yields are not
affected by the steep fall of the baryo-chemical potential. 
Since $\langle\Lambda\rangle \gg \langle\overline{\Lambda}\rangle$ 
and $\langle{\rm K}^+\rangle \gg \langle{\rm K}^-\rangle$ 
in the interval from top AGS to low SPS energies, the $\Lambda$ 
hyperons and K$^+$ carry a major fraction of the  overall $s$ and
$\overline{s}$ quark production. Thus $s + \overline{s}$ production, 
relative to $u + \overline{u} + d +\overline{d}$ production 
(as captured mostly in the pion yield) must reach a maximum within 
the interval 
$\sqrt{s} \approx$ 5~GeV (top AGS energy) and $\sqrt{s}$ = 8.7~GeV 
(the lowest SPS energy covered in the present study). 
As the corresponding p-p ratios vary much less over this energy range 
we finally conclude that the relative "strangeness enhancement" 
in A-A collisions reaches a maximum within this range.

The observed $\sqrt{s}$-dependence is confronted in Figs.~\ref{fig4} 
and~\ref{fig5} 
with predictions from the statistical hadronization 
model~\cite{pbm02} and from the microscopic transport models 
UrQMD~\cite{Web302,Web202} and HSD~\cite{Web302,Cass02}.
The former employs a ($T, \mu_B$) relation derived from a wide 
body of hadron production data~\cite{pbm02}.
The hadron gas model closely reproduces the 
$\langle\Lambda\rangle/\langle\pi\rangle$ ratio, but it underpredicts 
the $\langle\overline{\Lambda}\rangle/\langle\pi\rangle$ measurements. 
The transport models also predict the main trend of the energy dependence
of the ratios. However, they do not provide a quantitative description.

In summary, we have presented evidence for a relative strangeness
enhancement maximum within the interval $5 \le \sqrt{s} \le 8$ GeV, 
as inferred both from the present hyperon
data and from our previous kaon data~\cite{KPI02}. Upcoming analysis 
of data gathered at the SPS inside this interval will decide as
to whether the relatively smooth maximum implied by the ($T, \mu_B$) 
relation assumed in the statistical hadronization model~\cite{pbm02}
captures the detailed features of that strangeness peak. First
such K$^+/\pi^+$ results obtained at $\sqrt{s}$ = 7 GeV~\cite{Fri03} 
seem to rather indicate a sharp peak as was predicted for the onset 
of deconfinement, e.g. in Ref.~\cite{GazGor99}.

\begin{acknowledgments}
Acknowledgements: This work was supported by 
the US Department of Energy Grant DE-FG03-97/ER41020/A000,  
the Bundesministerium f\"ur Bildung und Forschung, Germany, 
the Polish State Committee for Scientific Research (2 P03B 130 23, 
SPB/CERN/P-03/Dz 446/2002-2004, 2 P03B 02418, 2 P03B 04123),  
the Hungarian Scientific Research Fund (T032648, T043514, T32293 
and F034707)
the Polish-German Foundation,
and the Korea Research Foundation Grant (KRF-2003-041-C00088).
\end{acknowledgments}
\bibliography{draft10.bbl}
\end{document}